\documentclass[twocolumn,epjc3,final]{svjour3}
\RequirePackage[T1]{fontenc}

\smartqed  

\RequirePackage{graphicx}
\RequirePackage{mathptmx}      
\RequirePackage{flushend}
\RequirePackage[numbers,sort&compress]{natbib}

\usepackage{amsmath, txfonts,multirow}
\usepackage{graphicx,bm,booktabs}
\usepackage{color}
\usepackage[bookmarks=true,bookmarksopen=false,plainpages=false,breaklinks=true,
   bookmarksnumbered=true,hypertexnames=false,
   filecolor=blue,urlcolor=three,menucolor=three,
   linkcolor=three,citecolor=blueone, colorlinks,
   anchorcolor=blue,runcolor=pink,frenchlinks=red
   pdfstartview=FitH,pdftitle=title,%
   pdfauthor=author]{hyperref}
   
\allowdisplaybreaks[4]

\definecolor{cover}{rgb}{0.77,0.87,0.88}
\definecolor{blueone}{rgb}{0.1,0.1,.7}
\definecolor{citec}{rgb}{0.14,0.47,0.09}
\definecolor{two}{rgb}{0.0,0.5,0.}
\definecolor{three}{rgb}{.5,.1,0.15}

\journalname{Eur. Phys. J. C}

\begin{document}

\title{Molecular states  from $D^{(*)}\bar{D}^{(*)}/B^{(*)}\bar{B}^{(*)}$ and $D^{(*)}D^{(*)}/\bar{B}^{(*)}\bar{B}^{(*)}$ interactions}
\author{Zuo-Ming Ding, Han-Yu Jiang, Jun He\thanksref{e1}}
\thankstext{e1}{Corresponding author: junhe@njnu.edu.cn}
\institute{Department of  Physics and Institute of Theoretical Physics, Nanjing Normal University,
Nanjing 210097, China
}
\date{Received: date / Revised version: date}

\maketitle
\begin{abstract}

In this work, we preform a systematic investigation about  hidden heavy and  doubly heavy molecular states from the $D^{(*)}\bar{D}^{(*)}/B^{(*)}\bar{B}^{(*)}$ and $D^{(*)}D^{(*)}/\bar{B}^{(*)}\bar{B}^{(*)}$ interactions in the quasipotential Bethe-Salpeter equation (qBSE) approach. With the help of  Lagrangians with heavy quark  and chiral symmetries,  interaction potentials are constructed within the one-boson-exchange model in which we include the $\pi$, $\eta$, $\rho$, $\omega$ and $\sigma$ exchanges, as well as $J/\psi$ or $\Upsilon$ exchange.  Possible bound states from the interactions considered are searched for as the pole of scattering amplitude. The results suggest that  experimentally observed states, $Z_c(3900)$,  $Z_c(4020)$, $Z_b(10610)$,  and $Z_b(10650)$,  can be related to the $D\bar{D}^{*}$,  $D^*\bar{D}^{*}$, $B\bar{B}^{*}$,  and $B^*\bar{B}^{*}$  interactions with quantum numbers $I^G(J^P)=1^+(1^{+})$, respectively. The $D\bar{D}^{*}$  interaction  is also attractive enough to produce a pole with $0^+(0^+)$ which is related  to the $X(3872)$.  Within the same theoretical frame, the existence of $D\bar{D}$ and $B\bar{B}$ molecular states with $0(0^+)$ are predicted. The possible $D^*\bar{D}^*$ molecular states with $0(0^+, 1^+, 2^+)$  and $1(0^+)$ and their bottom partners are also suggested by the calculation. In the doubly heavy sector, no bound state is produced from the $DD/\bar{B}\bar{B}$ interaction  while a bound state is found with $0(1^+)$ from $DD^*/\bar{B}\bar{B}^*$ interaction. The $D^*D^*/\bar{B}^*\bar{B}^*$ interaction produces three molecular states with $0(1^+)$, $0(2^+)$ and $1(2^+)$. 

\end{abstract}

\section{Introduction}

In recent years, more and more XYZ particles were observed in experiment~\cite{Zyla:2020zbs,Olsen:2017bmm}.  Such particles inspire  many theoretical efforts to interpret their origin and internal structure~\cite{Chen:2016qju,Guo:2017jvc,Ali:2017jda,Dong:2017gaw,Guo:2019twa}.  Due to  lack of  high-precision experimental data  and  difficulty to make a precise theoretical calculation, the origin of XYZ particles is still in debate.  Besides  non-particle interpretations, such as  cusp effect and anomalous triangle singularity~\cite{Guo:2019twa,Bugg:2011jr,Bugg:2004rk},  molecular state  and compact multiquark  are the most important pictures to explain such particles~\cite{Chen:2016qju,Guo:2017jvc,Ali:2017jda,Dong:2017gaw}.  The molecular state is a loosely bound states of two or more hadrons. It has an obvious characteristic that its mass is close to a threshold of hadrons.  Interestingly, many XYZ particles were observed near a threshold~\cite{Zyla:2020zbs}.  Such phenomenon  is the physical basis of  molecular state interpretation of the XYZ particle, as well as the cusp effect and anomalous triangle singularity.  The characteristic suggests that we should investigate possible molecular states in every threshold if the molecular state picture is true.   In the current work, we   focus on two important thresholds,  $D^{(*)}D^{(*)}$ and $B^{(*)}B^{(*)}$, where many important XYZ particles were observed, such as $X(3872)$, $Z_c(3900)$, $Z_c(4020)$, $Z_b(10610)$, and $Z_b(10650)$~\cite{Zyla:2020zbs,Chen:2016qju,Guo:2017jvc,Guo:2019twa,Dong:2017gaw}. 

The charmonium-like and bottonnmium-like states near the  $D^{(*)}D^{(*)}/B^{(*)}B^{(*)}$ thresholds has been widely studied in the literature. The $X(3872)$ is the first observed XYZ particle, and was explained as a $D\bar{D}^*$ molecular state very soon  after observation because it almost stands on the $D\bar{D}^*$ threshold~\cite{Choi:2003ue,Tornqvist:2004qy,Swanson:2003tb}.  Though later studies suggest that the $c\bar{c}$ component may be important to explain its property, in most models, the $D\bar{D}^*$ component is still very important~\cite{Close:2003sg,Barnes:2003vb,Kalashnikova:2005ui}.  The $Z_c(3900)$ and $Z_c(4020)$ are another two important XYZ particles, which were interpreted as  isovector $D\bar{D}^*$ molecular states in the literature~\cite{Ablikim:2013mio,Wang:2013cya, Liu:2013dau,Wang:2013daa,Zhang:2013aoa,Aceti:2014uea,He:2014nya,He:2013nwa}. However, some lattice calculations do not support such assignment~\cite{Prelovsek:2014swa,Chen:2014afa}. In the bottom sector, the $Z_b(10610)$ and $Z_b(10650)$ are good candidates of $B\bar{B}^*$ and $B^*\bar{B}^*$ molecular states~\cite{Garmash:2015rfd,Sun:2011uh,Zhang:2011jja,Wang:2014gwa,Dias:2014pva,Molina:2009ct,Ozpineci:2013zas}. Generally speaking, the observed charmonium-like and bottonnmium-like states construct a quite good pattern for the hidden-heavy (hidden-charm or hidden-bottom) molecular states near $D^{(*)}D^{(*)}/B^{(*)}B^{(*)}$ thresholds with some exceptions.  

To confirm the molecular picture for the states near $D^{(*)}D^{(*)}/B^{(*)}B^{(*)}$ thresholds,  more high-precision experimental data and theoretical predictions are required.  In Ref.~\cite{Gamermann:2006nm}, a $D\bar{D}$ bound state was proposed and experimentally searched  by the Belle Collaboration~\cite{Chilikin:2017evr}.  The existence of $D^{(*)}\bar{D}^{(*)}/B^{(*)}\bar{B}^{(*)}$  molecular states with $0^+$ and $2^+$ are also discussed in the literature~\cite{Nieves:2012tt,Baru:2017gwo, Baru:2019xnh}. In the other side, if we find more states near these thresholds,  the molecular state interpretation of observed XYZ particles is also supported.  All possible charmonium-like and bottonnmium-like states  near $D^{(*)}D^{(*)}/B^{(*)}B^{(*)}$ have been searched in the experiment.  If we want to find more states, it is a good way to study the doubly heavy sector, which  thresholds are almost the same as the hidden heavy ones. Theoretically, the double heavy sector can be obtained by replacing one constituent particle by its antiparticle. The similarity between particle and antiparticle guarantees the similarity of two interactions. If the molecular states in the hidden heavy sector exist,  the existence of the doubly heavy  states predicted in the same model can be expected. 

In the literature, the studies about doubly heavy molecular states are relatively scarce due to difficulty in experimental production. Barnes suggested that  isoscalar $BB^*$ interaction is attractive and a vector state with spin parity $1^+$ maybe exist by solving the Sch\"odinger equation~\cite{Barnes:1999hs}.  The $DD^*$ and $\bar{B}\bar{B}^*$ interactions were studied within the framework of heavy meson chiral effective field theory~\cite{Wang:2018atz,Xu:2017tsr}.  An isovector bound state was found from the $DD^*$ interaction~\cite{Xu:2017tsr}.  In Ref.~\cite{Li:2012ss}, the one-boson-exchange model was also applied to study the possible double-heavy molecular states.  Recently, the LHCb reported the observation of  doubly charmed baryon state $\Xi_{cc}^{++}$~\cite{Aaij:2017ueg}, which raises the hope of experimental observation of the doubly heavy molecular states.   In Ref.~\cite{Yu:2019sxx},  possible $\bar{B}^{(*)}\bar{B}^{(*)}$ molecular states were studied in a constituent interchange model .  

In our previous works, the molecular states from $D^{(*)}\bar{D}^{(*)}/{B}^{(*)}\bar{B}^{(*)}$ interactions were studied in different approaches,  non-relativistic Schr\"odinger equation and the qBSE approach~\cite{Sun:2011uh,He:2014nya,Sun:2012zzd,He:2015mja,He:2017lhy}.  In the current work, we will perform a systematic study of possible molecular states produced from the $D^{(*)}\bar{D}^{(*)}/{B}^{(*)}\bar{B}^{(*)}$ and $D^{(*)}D^{(*)}/\bar{B}^{(*)}\bar{B}^{(*)}$ interactions in the qBSE approach combined with the one-boson-exchange model. The hidden heavy molecular states will be studied and related to the states observed in experiment. In the same theoretical frame,  one of two constituent particles will be replaced by its antiparticle,  and the doubly heavy molecular state can  be studied. 

This article is organized as follows. After introduction, we present the details of theoretical frame in section~\ref{Sec: Formalism}, which include flavor wave functions, effective Lagrangians, construction of potential and a brief introduction of the qBSE approach.  The numerical results for the states produced from the interaction considered will be given in Section~\ref{Sec: Results}. Finally,  article ends with a summary in section~\ref{Sec: Summary}.

\section{Theoretical frame} \label{Sec: Formalism}

First, we should construct   flavor wave functions with definite
isospin under SU(3) symmetry. For the  $D\bar{D}^*$ states, we have~\cite{Sun:2011uh},
\begin{align}
|X_{D\bar{D}^*}^0\rangle_{I=0}&=\frac{1}{2}\Big[\big(|D^{*+}D^-\rangle+|D^{*0}\bar{D}^0\rangle\big)+c\big(|D^+D^{*-}\rangle+|D^0\bar{D}^{*0}\rangle\big)\Big], \nonumber\\
|X_{D\bar{D}^*}^0\rangle_{I=1}&=\frac{1}{2}\Big[\big(|D^{*+}D^-\rangle-|D^{*0}\bar{D}^0\rangle\big)+c\big(|D^+D^{*-}\rangle-|D^0\bar{D}^{*0}\rangle\big)\Big],\nonumber\\
|X_{D\bar{D}^*}^+\rangle_{I=1}&=\frac{1}{\sqrt{2}}\big(|D^{*+}\bar{D}^0\rangle+c|D^+\bar{D}^{*0}\rangle\big),\nonumber\\
|X_{D\bar{D}^*}^-\rangle_{I=1}&=\frac{1}{\sqrt{2}}\big(|D^{*-}\bar{D}^0\rangle+c|D^-\bar{D}^{*0}\rangle\big),
 \label{Eq: wf1}
\end{align}
where $c=\pm$ corresponds to $C$ parity $C=\mp$ respectively. For the isovector state, the $C$ parity can not be defined, so we will use the $G$ parity instead. For the $D\bar{D}$ states, the wave functions can be constructed as
\begin{align}
|X_{D\bar{D}}^0\rangle_{I=0}&=\frac{1}{\sqrt{2}}\left(|D^{+}\bar{D}^{-}\rangle+|D^0\bar{D}^{0}\rangle\right),\nonumber\\
|X_{D\bar{D}}^0\rangle_{I=1}&=\frac{1}{\sqrt{2}}\left(|D^{+}\bar{D}^{-}\rangle-|D^0\bar{D}^{0}\rangle\right),\nonumber\\
|X_{D\bar{D}}^+\rangle_{I=1}&=|D^{+}\bar{D}^{0}\rangle,\nonumber\\
|X_{D\bar{D}}^-\rangle_{I=1}&=|D^{-}\bar{D}^{0}\rangle.
 \label{Eq: wf2}
\end{align}
Since $G$ parity is definite for these states,  there is no additional $c$ in the definition of  wave functions. The wave functions of $D^*\bar{D}^*$ states have analogous forms as   $D\bar{D}$ states. 

The wave functions for $D^{(*)}D^{(*)}$ states can be also constructed under  SU(3) symmetry. Since doubly charmed states have no $G$ parity, their wave functions are in forms analogous to the $D\bar{D}$ states as,
\begin{align}
|X_{D^{(*)}{D}^{(*)}}^+\rangle_{I=0}&=\frac{1}{\sqrt{2}}\Big(|D^{(*)+}D^{(*)0}\rangle-|D^{(*)0}{D}^{(*)+}\rangle\Big), \nonumber\\
|X_{D^{(*)}{D}^{(*)}}^+\rangle_{I=1}&=\frac{1}{\sqrt{2}}\Big(|D^{(*)+}D^{(*)0}\rangle+|D^{(*)0}{D}^{(*)+}\rangle\Big), \nonumber\\
|X_{D^{(*)}{D}^{(*)}}^{0}\rangle_{I=1}&=|D^{(*)0}D^{(*)0}\rangle,\nonumber\\
|X_{D^{(*)}{D}^{(*)}}^{++}\rangle_{I=1}&=|D^{(*)+}\bar{D}^{(*)+}\rangle.
 \label{Eq: wf3}
\end{align}
The wave functions of the hidden bottom and doubly bottom states can be obtained analogously. 

To obtain the interaction between the heavy mesons, we adopt the one-boson-exchange model where the pseudoscalar meson ($\mathbb{P}=\pi$ and  $\eta$), vector meson ($\mathbb{V}=\rho$ and $\omega$), and scalar meson ($\sigma$) are considered.  According to the chiral
symmetry and heavy quark limit, the Lagrangian for heavy mesons interacting with light mesons reads\cite{Cheng:1992xi,Yan:1992gz,Wise:1992hn,Burdman:1992gh,Casalbuoni:1996pg}
\begin{align}
\mathcal{L}_{\mathcal{P}^*\mathcal{P}^*\mathbb{P}} &=
-i\frac{2g}{f_\pi}\varepsilon_{\alpha\mu\nu\lambda}
v^\alpha\mathcal{P}^{*\mu}_{b}{\mathcal{P}}^{*\lambda\dag}_{a}
\partial^\nu{}\mathbb{P}_{ba}\nonumber\\
&+i \frac{2g}{f_\pi}\varepsilon_{\alpha\mu\nu\lambda}
v^\alpha\widetilde{\mathcal{P}}^{*\mu\dag}_{a}\widetilde{\mathcal{P}}^{*\lambda}_{b}
\partial^\nu{}\mathbb{P}_{ab},\nonumber\\
\mathcal{L}_{\mathcal{P}^*\mathcal{P}\mathbb{P}} &=-
\frac{2g}{f_\pi}(\mathcal{P}^{}_b\mathcal{P}^{*\dag}_{a\lambda}+
\mathcal{P}^{*}_{b\lambda}\mathcal{P}^{\dag}_{a})\partial^\lambda{}
\mathbb{P}_{ba}\nonumber\\
&+\frac{2g}{f_\pi}(\widetilde{\mathcal{P}}^{*\dag}_{a\lambda}\widetilde{\mathcal{P}}_b+
\widetilde{\mathcal{P}}^{\dag}_{a}\widetilde{\mathcal{P}}^{*}_{b\lambda})\partial^\lambda{}\mathbb{P}_{ab}.
\nonumber\\
  \mathcal{L}_{\mathcal{PP}\mathbb{V}}
  &= -\sqrt{2}\beta{}g_V\mathcal{P}^{}_b\mathcal{P}_a^{\dag}
  v\cdot\mathbb{V}_{ba}
 +\sqrt{2}\beta{}g_V\widetilde{\mathcal{P}}^{\dag}_a
  \widetilde{\mathcal{P}}^{}_b
  v\cdot\mathbb{V}_{ab},\nonumber\\
  \mathcal{L}_{\mathcal{P}^*\mathcal{P}\mathbb{V}}
  &=- 2\sqrt{2}\lambda{}g_V v^\lambda\varepsilon_{\lambda\mu\alpha\beta}
  (\mathcal{P}^{}_b\mathcal{P}^{*\mu\dag}_a +
  \mathcal{P}_b^{*\mu}\mathcal{P}^{\dag}_a)
  (\partial^\alpha{}\mathbb{V}^\beta)_{ba}\nonumber\\
&-  2\sqrt{2}\lambda{}g_V
v^\lambda\varepsilon_{\lambda\mu\alpha\beta}
(\widetilde{\mathcal{P}}^{*\mu\dag}_a\widetilde{\mathcal{P}}^{}_b
+
\widetilde{\mathcal{P}}^{\dag}_a\widetilde{\mathcal{P}}_b^{*\mu})
  (\partial^\alpha{}\mathbb{V}^\beta)_{ab},\nonumber\\
  \mathcal{L}_{\mathcal{P}^*\mathcal{P}^*\mathbb{V}}
  &= \sqrt{2}\beta{}g_V \mathcal{P}_b^{*}\cdot\mathcal{P}^{*\dag}_a
  v\cdot\mathbb{V}_{ba}\nonumber\\
  &-i2\sqrt{2}\lambda{}g_V\mathcal{P}^{*\mu}_b\mathcal{P}^{*\nu\dag}_a
  (\partial_\mu{}
  \mathbb{V}_\nu - \partial_\nu{}\mathbb{V}_\mu)_{ba}\nonumber\\
  &-\sqrt{2}\beta g_V
  \widetilde{\mathcal{P}}^{*\dag}_a\widetilde{\mathcal{P}}_b^{*}
  v\cdot\mathbb{V}_{ab}\nonumber\\
  &-i2\sqrt{2}\lambda{}g_V\widetilde{\mathcal{P}}^{*\mu\dag}_a\widetilde{\mathcal{P}}^{*\nu}_b(\partial_\mu{}
  \mathbb{V}_\nu - \partial_\nu{}\mathbb{V}_\mu)_{ab}.
\nonumber\\
  \mathcal{L}_{\mathcal{PP}\mathbb{V}}
  &= -\sqrt{2}\beta{}g_V\mathcal{P}^{}_b\mathcal{P}_a^{\dag}
  v\cdot\mathbb{V}_{ba}
 +\sqrt{2}\beta{}g_V\widetilde{\mathcal{P}}^{\dag}_a
  \widetilde{\mathcal{P}}^{}_b
  v\cdot\mathbb{V}_{ab},\nonumber\\
  \mathcal{L}_{\mathcal{P}^*\mathcal{P}\mathbb{V}}
  &=- 2\sqrt{2}\lambda{}g_V v^\lambda\varepsilon_{\lambda\mu\alpha\beta}
  (\mathcal{P}^{}_b\mathcal{P}^{*\mu\dag}_a +
  \mathcal{P}_b^{*\mu}\mathcal{P}^{\dag}_a)
  (\partial^\alpha{}\mathbb{V}^\beta)_{ba}\nonumber\\
&-  2\sqrt{2}\lambda{}g_V
v^\lambda\varepsilon_{\lambda\mu\alpha\beta}
(\widetilde{\mathcal{P}}^{*\mu\dag}_a\widetilde{\mathcal{P}}^{}_b
+
\widetilde{\mathcal{P}}^{\dag}_a\widetilde{\mathcal{P}}_b^{*\mu})
  (\partial^\alpha{}\mathbb{V}^\beta)_{ab},\nonumber\\
  \mathcal{L}_{\mathcal{P}^*\mathcal{P}^*\mathbb{V}}
  &= \sqrt{2}\beta{}g_V \mathcal{P}_b^{*}\cdot\mathcal{P}^{*\dag}_a
  v\cdot\mathbb{V}_{ba}\nonumber\\
  &-i2\sqrt{2}\lambda{}g_V\mathcal{P}^{*\mu}_b\mathcal{P}^{*\nu\dag}_a
  (\partial_\mu{}
  \mathbb{V}_\nu - \partial_\nu{}\mathbb{V}_\mu)_{ba}\nonumber\\
  &-\sqrt{2}\beta g_V
  \widetilde{\mathcal{P}}^{*\dag}_a\widetilde{\mathcal{P}}_b^{*}
  v\cdot\mathbb{V}_{ab}\nonumber\\
  &-i2\sqrt{2}\lambda{}g_V\widetilde{\mathcal{P}}^{*\mu\dag}_a\widetilde{\mathcal{P}}^{*\nu}_b(\partial_\mu{}
  \mathbb{V}_\nu - \partial_\nu{}\mathbb{V}_\mu)_{ab}.
\nonumber\\
  \mathcal{L}_{\mathcal{PP}\sigma}
  &= -2g_s\mathcal{P}^{}_b\mathcal{P}^{\dag}_b\sigma
 -2g_s\widetilde{\mathcal{P}}^{}_b\widetilde{\mathcal{P}}^{\dag}_b\sigma,\nonumber\\
  \mathcal{L}_{\mathcal{P}^*\mathcal{P}^*\sigma}
  &= 2g_s\mathcal{P}^{*}_b\cdot{}\mathcal{P}^{*\dag}_b\sigma
 +2g_s\widetilde{\mathcal{P}}^{*}_b\cdot{}\widetilde{\mathcal{P}}^{*\dag}_b\sigma,\label{Eq:L}
\end{align} 
where  the velocity $v$ should be replaced by $i\overleftrightarrow{\partial}/2\sqrt{m_im_f}$ with the $m_{i,f}$ being the mass of the initial or final heavy meson. 
${\mathcal{P}}^{(*)T} =(D^{(*)0},D^{(*)+},D_s^{(*)+})$ or
$(B^{(*)-},\bar{B}^{(*)0},\bar{B}_s^{(*)0})$, and 
 satisfy the normalization relations $\langle
0|{\mathcal{P}}|{Q}\bar{q}(0^-)\rangle
=\sqrt{M_\mathcal{P}}$ and $\langle
0|{{\mathcal{P}}}^*_\mu|{Q}\bar{q}(1^-)\rangle=
\epsilon_\mu\sqrt{M_{\mathcal{P}^*}}$. 
The $\mathbb
P$ and $\mathbb V$ are the pseudoscalar and vector matrices
\begin{equation}
    {\mathbb P}=\left(\begin{array}{ccc}
        \frac{\sqrt{3}\pi^0+\eta}{\sqrt{6}}&\pi^+&K^+\\
        \pi^-&\frac{-\sqrt{3}\pi^0+\eta}{\sqrt{6}}&K^0\\
        K^-&\bar{K}^0&-\frac{2\eta}{\sqrt{6}}
\end{array}\right),
\mathbb{V}=\left(\begin{array}{ccc}
\frac{\rho^0+\omega}{\sqrt{2}}&\rho^{+}&K^{*+}\\
\rho^{-}&\frac{-\rho^{0}+\omega}{\sqrt{2}}&K^{*0}\\
K^{*-}&\bar{K}^{*0}&\phi
\end{array}\right).\label{MPV}
\end{equation}
The parameters involved here were determined in the literature as $g=0.59$, $\beta=0.9$, $\lambda=0.56$ GeV$^{-1}$, and $g_s=0.76$ with $g_V=5.9$ and  $f_\pi=132$ MeV~\cite{Falk:1992cx,Isola:2003fh,Liu:2008tn,Chen:2019asm}.

In Refs.~\cite{Aceti:2014uea,He:2015mja}, contribution from the $J/\psi$ exchange is found important in the $D\bar{D}^*$ interaction to produce the $Z_c(3900)$ observed at BESIII.  In the current work, we also consider such exchange with the couplings of heavy-light charmed mesons to $J/\psi$, which are written with the help of  heavy quark effective theory as \cite{Casalbuoni:1996pg,Oh:2000qr},
\begin{eqnarray}
	{\cal L}_{D^*_{(s)}\bar{D}^*_{(s)}J/\psi}&=&-ig_{D^*_{(s)}D^*_{(s)}\psi}\big[\psi \cdot \bar{D}^*\overleftrightarrow{\partial}\cdot D^*\nonumber\\
&-&
\psi^\mu \bar D^* \cdot\overleftrightarrow{\partial}^\mu {D}^* +
\psi^\mu \bar{D}^*\cdot\overleftrightarrow{\partial} D^{*\mu} ) \big], \nonumber \\
{\cal L}_{D_{(s)}^*\bar{D}_{(s)}J/\psi}&=&
g_{D^*_{(s)}D_{(s)}\psi} \,  \, \epsilon_{\beta \mu \alpha \tau}
\partial^\beta \psi^\mu (\bar{D}
\overleftrightarrow{\partial}^\tau D^{* \alpha}+\bar{D}^{* \alpha}
\overleftrightarrow{\partial}^\tau D) \label{matrix3}, \nonumber \\
{\cal L}_{D_{(s)} \bar{D}_{(s)}J/\psi} &=&
ig_{D_{(s)}D_{(s)}\psi} \psi \cdot
\bar{D}\overleftrightarrow{\partial}D,
\end{eqnarray}
where the couplings
are related to a single parameter $g_2$ as
\begin{eqnarray}
\frac{g_{D^*D^*\psi} }{m_{D^*}}= \frac{g_{D_{(s)}D_{(s)}\psi}}{m_D}= g_{D^*_{(s)}D_{(s)}\psi}= 2 g_2 \sqrt{m_\psi },\label{Eq: para}
\end{eqnarray}
with $g_2={\sqrt{m_\psi}}/({2m_Df_\psi})$ and $f_\psi=405$ MeV.  For the bottom mesons,  analogous Lagrangians can be obtained under the heavy quark symmetry for  $\Upsilon$ exchange. The parameters can be obtained by similar relation in Eq.~(\ref{Eq: para}) by replacing the mass by these of bottom mesons and $f_\Upsilon=715$ MeV~\cite{Li:2012as}. 

With the vertices  obtained from the former Lagrangians, the potential  interaction can be constructed in the one-boson-exchange model as~\cite{He:2019ify}
\begin{equation}%
{\cal V}_{\mathbb{P},\sigma}=I^{(d,c)}_i\Gamma_1\Gamma_2 P_{\mathbb{P},\sigma}f(q^2),\ \
{\cal V}_{\mathbb{V}}=I^{(d,c)}_i\Gamma_{1\mu}\Gamma_{2\nu}  P^{\mu\nu}_{\mathbb{V}}f(q^2),\label{V}
\end{equation}
where the propagators are defined as usual as
\begin{equation}%
P_{\mathbb{P},\sigma}= \frac{i}{q^2-m_{\mathbb{P},\sigma}^2},\ \
P^{\mu\nu}_\mathbb{V}=i\frac{-g^{\mu\nu}+q^\mu q^\nu/m^2_{\mathbb{V}}}{q^2-m_\mathbb{V}^2}.
\end{equation}
We introduce a form factor $f(q^2)=\Lambda_e^2/(q^2-\Lambda_e^2)$  to reflect the off-shell effect of exchanged meson 
with $q$ being the momentum of the exchanged  meson as in Refs.~\cite{He:2015mja,Gross:1991pm}.

In our approach, we collect the coefficients for the interaction of a state as a flavor factor.  As discussed in Ref.~\cite{He:2014nya}, for the hidden heavy state, the cross diagram  appears in the $D\bar{D}^*$ and $B\bar{B}^*$ cases due to the coupling between the two parts with and without $c$ as shown in Eq.~(\ref{Eq: wf1}).  Obviously, for the states,  $D\bar{D}/B\bar{B}$ nad $D^*\bar{D}^*/B^*\bar{B}^*$, there is no cross diagram.  For the doubly heavy states,  there is no cross diagram from the coupling in  the $D\bar{D}^*$ and $B\bar{B}^*$ cases. However, in such cases, the $u$ channel is allowed, which will provide cross diagrams.   In Table~\ref{flavor factor},  flavor factors $I^d_i$ and $I^c_i$ of certain meson exchange $i$ of certain interaction are listed for direct and cross diagrams, respectively.
\renewcommand\tabcolsep{0.108cm}
\renewcommand{\arraystretch}{1.5}
\begin{table}[h!]
\begin{center}
\caption{The isospin factors $I_i^d$ and $I_i^c$ for direct and cross diagrams and different exchange mesons. 
\label{flavor factor}}
\begin{tabular}{c|cccccc|ccccccc}\bottomrule[2pt]
 & \multicolumn{6}{c|}{$I_i^d$}& \multicolumn{6}{c}{ $I_i^c$}\\\hline
&$\pi$&$\eta$  &$\rho$ &$\omega$&$\sigma$ &$J/\psi$ &$\pi$&$\eta$  &$\rho$ &$\omega$&$\sigma$ &$J/\psi$ \\\hline
$[\mathcal{P}\bar{\mathcal{P}}]^T$&$--$&$--$&$-\frac{1}{2}$ &$\frac{1}{2}$ & $1$ &$1$
&$--$&$--$&$--$ &$--$ & $--$ &$--$\\
$[\mathcal{P}\bar{\mathcal{P}}]^S$&$--$&$--$&$\frac{3}{2}$&$\frac{1}{2}$& $1$ &$1$
&$--$&$--$&$--$ &$--$ & $--$ &$--$\\
$[\mathcal{P}\bar{\mathcal{P}}^{*}]^T$&$--$&$--$&$-\frac{1}{2}$ &$\frac{1}{2}$ &$1$ &$1$
&$-\frac{c}{2}$&$\frac{c}{6}$&$-\frac{c}{2}$ &$\frac{c}{2}$ & $--$ &$c$\\
$[\mathcal{P}\bar{\mathcal{P}}^{*}]^S$&$--$&$--$&$\frac{3}{2}$&$\frac{1}{2}$&$1$ &$1$
&$\frac{3}{2}c$&$\frac{c}{6}$&$\frac{3}{2}c$ &$\frac{c}{2}$ & $--$ &$c$\\
$[\mathcal{P}^{*}\bar{\mathcal{P}}^{*}]^T$&$-\frac{1}{2}$&$\frac{1}{6}$&$-\frac{1}{2}$ &$\frac{1}{2}$ & $1$ &$1$
&$--$&$--$&$--$ &$--$ & $--$ &$--$\\
$[\mathcal{P}^{*}\bar{\mathcal{P}}^{*}]^S$&$\frac{3}{2}$&$\frac{1}{6}$&$\frac{3}{2}$&$\frac{1}{2}$& $1$ &$1$
&$--$&$--$&$--$ &$--$ & $--$ &$--$\\
\bottomrule[1.pt]
$[\mathcal{P}\mathcal{P}]^T$&$--$&$--$&$\frac{1}{2}$ &$\frac{1}{2}$ & $1$ &$1$
&$--$&$--$&$\frac{1}{2}$ &$\frac{1}{2}$ & $1$ &$1$\\
$[\mathcal{P}\mathcal{P}]^S$&$--$&$--$&$-\frac{3}{2}$&$\frac{1}{2}$& $1$ &$1$
&$--$&$--$&$\frac{3}{2}$ &$-\frac{1}{2}$ & $1$ &$1$\\
$[\mathcal{P}\mathcal{P}^{*}]^T$&$--$&$--$&$\frac{1}{2}$ &$\frac{1}{2}$ & $1$& $1$
&$\frac{1}{2}$&$\frac{1}{6}$&$\frac{1}{2}$ &$\frac{1}{2}$ & $--$ &$1$\\
$[\mathcal{P}\mathcal{P}^{*}]^S$&$--$&$--$&$-\frac{3}{2}$&$\frac{1}{2}$&$1$ &$1$
&$\frac{3}{2}$&$-\frac{1}{6}$&$\frac{3}{2}$ &$-\frac{1}{2}$ & $--$ &$1$\\
$[\mathcal{P}^{*}\mathcal{P}^{*}]^T$&$\frac{1}{2}$&$\frac{1}{6}$&$\frac{1}{2}$ &$\frac{1}{2}$ & $1$ &$1$
&$\frac{1}{2}$&$\frac{1}{6}$&$\frac{1}{2}$&$\frac{1}{2}$& $1$ &$1$\\
$[\mathcal{P}^{*}\mathcal{P}^{*}]^S$&$-\frac{3}{2}$&$\frac{1}{6}$&$-\frac{3}{2}$&$\frac{1}{2}$& $1$ &$1$
&$\frac{3}{2}$&$-\frac{1}{6}$&$\frac{3}{2}$&$-\frac{1}{2}$& $1$ &$1$\\
\toprule[2pt]
\end{tabular}
\end{center}
\end{table}

In the above, we construct the potential of the interactions considered in the current work. The scattering amplitude can be obtained with the qBSE~\cite{He:2014nya,He:2015mja,He:2017lhy,He:2015yva,He:2017aps}. After  the partial-wave decomposition,  the qBSE can be reduced to a 1-dimensional  equation with a spin-parity $J^P$ as~\cite{He:2015mja},
\begin{align}
i{\cal M}^{J^P}_{\lambda'\lambda}({\rm p}',{\rm p})
&=i{\cal V}^{J^P}_{\lambda',\lambda}({\rm p}',{\rm
p})+\sum_{\lambda''}\int\frac{{\rm
p}''^2d{\rm p}''}{(2\pi)^3}\nonumber\\
&\cdot
i{\cal V}^{J^P}_{\lambda'\lambda''}({\rm p}',{\rm p}'')
G_0({\rm p}'')i{\cal M}^{J^P}_{\lambda''\lambda}({\rm p}'',{\rm
p}),\quad\quad \label{Eq: BS_PWA}
\end{align}
where the ${\cal M}^{J^P}({\rm p}',{\rm p})$ is partial-wave scattering amplitude, and
the $G_0({\rm p}'')$ is reduced propagator with the spectator approximation~\cite{He:2015mja}.
The partial wave potential can be obtained from the potential in Eq.~(\ref{V}) as
\begin{align}
{\cal V}_{\lambda'\lambda}^{J^P}({\rm p}',{\rm p})
&=2\pi\int d\cos\theta
~[d^{J}_{\lambda\lambda'}(\theta)
{\cal V}_{\lambda'\lambda}({\bm p}',{\bm p})\nonumber\\
&+\eta d^{J}_{-\lambda\lambda'}(\theta)
{\cal V}_{\lambda'-\lambda}({\bm p}',{\bm p})],
\end{align}
where $\eta=PP_1P_2(-1)^{J-J_1-J_2}$ with $P$ and $J$ being parity and spin for system,  and two constituent heavy mesons. The initial and final relative momenta are chosen as ${\bm p}=(0,0,{\rm p})$  and ${\bm p}'=({\rm p}'\sin\theta,0,{\rm p}'\cos\theta)$. The $d^J_{\lambda\lambda'}(\theta)$ is the Wigner d-matrix.
An  exponential
regularization  was also introduced as  a form factor into the reduced propagator as $G_0({\rm p}'')\to G_0({\rm p}'')e^{-2(p''^2_l-m_l^2)^2/\Lambda_r^4}$~\cite{He:2015mja}.

\section{Numerical results}\label{Sec: Results}

With the input presented above, we can obtain the scattering amplitudes of the $D^{(*)}\bar{D}^{(*)}/B^{(*)}\bar{B}^{(*)}$ and $D^{(*)}D^{(*)}/\bar{B}^{(*)}\bar{B}^{(*)}$ interactions. Because we are only interesting in the pole of the scattering amplitude,  we only need to find the  position where $|1-V(W)G(W)|=0$ with $W$ equaling to  system energy $W$~\cite{He:2015mja,Gross:1991pm}.  In addition, we take two free parameters $\Lambda_e$ and $\Lambda_r$ as $\Lambda$ for simplification.

In Table~\ref{Tab: DD bound state} and Table~\ref{Tab: BB bound state}, we present the results in  charm and bottom sectors, respectively. In the current work, we only consider the spin parities which can be produced in S-wave, i.e., $0^+$ for $D\bar{D}/B\bar{B}$ and $DD/\bar{B}\bar{B}$ interactions, $1^+$ for  $D\bar{D}^*/B\bar{B}^*$ and $DD^*/\bar{B}\bar{B}^*$ interactions, and $0^+$, $1^+$ and $2^+$ for  $D^*\bar{D}^*/B^*\bar{B}^*$ and $D^*D^*/\bar{B}^*\bar{B}^*$ interactions. Here we scan the values of cutoff in a range smaller than 5 GeV and present results with some selected  values of cutoff $\Lambda$ if there is a bound state produced from the corresponding interaction. 
\renewcommand\tabcolsep{0.33cm}
\renewcommand{\arraystretch}{1.1}
\begin{table}[hbtp!]
\begin{center}
\caption{The binding energy $E_B$ of the bound states from the $D^{(*)}\bar{D}^{(*)}$ and $D^{(*)}{D}^{(*)}$ interactions with some selected values of cutoff $\Lambda$. The ``$--$" means that no bound state is found in the considered range of the cutoff, and blank means that the quantum number is forbidden. The $\Lambda$ and $W$ are in the units of GeV and MeV, respectively. 
\label{Tab: DD bound state}
\label{diagrams}}
	\begin{tabular}{c|rrrrrr}\bottomrule[2pt]
$$ & \multicolumn{2}{c}{$D\bar{D} (G=+)$}& \multicolumn{2}{c}{$D\bar{D}(G=-)$}& \multicolumn{2}{c}{$DD$}	 \\\hline
$I(J^{P})$   &  $\Lambda$ & $E_B$ &  $\Lambda$ & $E_B$ &  $\Lambda$ & $E_B$ \\\hline
$0(0^{+})$& 0.6  & 2.1 & $$ &  $$ & $--$ & $--$ \\
             & 0.8  & 10.7 & $$ &  $$ & $--$ & $--$ \\
              & 0.9  & 18.0 & $$ &  $$ & $--$ & $--$ \\
$1(0^{+})$&  &    & $3.5$  &0.4 & $--$ & $--$ \\
             &  &   & 3.9  &11.0& $--$ & $--$ \\
              & &   & 4.1  &22.2& $--$ & $--$ \\\bottomrule[1.pt]
 $$ & \multicolumn{2}{c}{$D\bar{D}^* (G=+)$}& \multicolumn{2}{c}{$D\bar{D}^*(G=-)$}& \multicolumn{2}{c}{$DD^*$}	 \\\hline 
$0(1^{+})$& 0.5 & 2.3 & 0.6 & 1.5 & 0.8 & 1.2\\
             & 0.7  & 13.4& 0.8& 12.0 & 1.0 & 12.7 \\
             & 0.8  & 23.3 & 0.9 & 22.7 & 1.1 & 27.9 \\
$1(1^{+})$& 1.9  &2.4& $--$  & $--$  & $--$  & $--$ \\
             & 2.0 &7.1& $--$  & $--$  & $--$  & $--$\\
             & 2.1 &17.1& $--$  & $--$  & $--$  & $--$ \\
           \bottomrule[1.pt]
 $$ & \multicolumn{2}{c}{$D^*\bar{D}^* (G=+)$}& \multicolumn{2}{c}{$D^*\bar{D}^*(G=-)$}& \multicolumn{2}{c}{$D^*D^*$}	 \\\hline 
$0(0^{+})$  & 0.8  & 0.8&  & & $--$  & $--$ \\
              & 1.4  & 14.0&  &  & $--$  & $--$ \\
              & 2.1  & 22.4 &  &  & $--$  & $--$ \\
$0(1^{+})$&  &  & 0.6& 1.0 & 0.9 & 1.0 \\
             &  & & 1.0 & 13.0 & 1.1 & 14.6 \\
             &   &  &1.2 & 23.5& 1.2 & 33.7 \\
$0(2^{+})$ & 0.5 & 2.4 &  &  & 0.7 & 0.3 \\
             & 0.7 & 14.0 &  &   & 1.0 & 15.2 \\
             & 0.8  & 25.6 & &  & 1.1 & 27.5 \\
$1(0^{+})$&  & & 1.8  & 1.9 & $--$  & $--$ \\
             &  &  & 2.1  & 11.5& $--$  & $--$ \\
             &   & & 2.3 &  23.1& $--$  & $--$ \\
$1(1^{+})$ &2.3 &2.8 &  && $--$  & $--$ \\
             & 2.6  &10.2&  &    & $--$  & $--$ \\
             & 2.8 & 20.0 &  & & $--$  & $--$ \\
$1(2^{+})$&  && 4.5  &1.6   & 1.75 & 1.0 \\
              & & & 4.9  & 10.4& 1.80 & 5.9 \\
              & && 5.0  &  18.2 & 1.90 & 41.4 \\
\toprule[2pt]
\end{tabular}
\end{center}

\end{table}

\renewcommand\tabcolsep{0.36cm}
\renewcommand{\arraystretch}{1.1}
\begin{table}[hbtp!]
\begin{center}
\caption{The binding energy $E_B$ of the bound states from the $B^{(*)}\bar{B}^{(*)}$ and $\bar{B}^{(*)}\bar{B}^{(*)}$ interactions with some selected values of cutoff $\Lambda$.  Other notations are the same as in Table~\ref{Tab: BB bound state}.
\label{Tab: BB bound state}
\label{diagrams}}
	\begin{tabular}{c|rrrrrr}\bottomrule[2pt]
$$ & \multicolumn{2}{c}{$B\bar{B} (G=+)$}& \multicolumn{2}{c}{$B\bar{B}(G=-)$}& \multicolumn{2}{c}{$\bar{B}\bar{B}$}	 \\\hline
$I(J^{P})$   &  $\Lambda$ & $E_B$ &  $\Lambda$ & $E_B$ &  $\Lambda$ & $E_B$ \\\hline
$0(0^{+})$& 0.3  & 1.1 &  & & $--$ &  $--$ \\
             & 0.5  & 9.2 & &  & $--$ &  $--$ \\
             & 0.6  & 17.2 &  &  & $--$ &  $--$ \\
$1(0^{+})$&  &  & 3.5  &1.7& $--$ &  $--$ \\
              & &   & 4.0  &11.1 & $--$ &  $--$ \\
             &  &  & 4.3   &22.9 & $--$ &  $--$ \\\bottomrule[1.pt]
 $$ & \multicolumn{2}{c}{$B\bar{B}^* (G=+)$}& \multicolumn{2}{c}{$B\bar{B}^*(G=-)$}& \multicolumn{2}{c}{$\bar{B}\bar{B}^*$}	 \\\hline 
$0(1^{+})$& 0.3 &1.5 & 1.2 & 2.2 & 0.4 & 1.0 \\
             & 0.5  & 11.9 & 1.4 & 12.6 &0.5& 9.3 \\
             & 0.6  & 21.4 & 1.5 & 22.0 & 0.7 & 18.2 \\
$1(1^{+})$& 1.6  &2.4& $--$ &  $--$ & $--$ &  $--$ \\
             & 1.9 & 10.9& $--$ &  $--$ & $--$ &  $--$ \\
             & 2.1  &22.7& $--$ &  $--$  & $--$ &  $--$ \\
            \bottomrule[1.pt]
 $$ & \multicolumn{2}{c}{$B^*\bar{B}^* (G=+)$}& \multicolumn{2}{c}{$B^*\bar{B}^*(G=-)$}& \multicolumn{2}{c}{$\bar{B}^*\bar{B}^*$}	 \\\hline 
$0(0^{+})$& 0.4  & 1.3 &  &   & $--$ &  $--$ \\
             & 0.8  & 9.5 &  &  & $--$ &  $--$ \\
             & 1.2  & 20.9 &  &   & $--$ &  $--$ \\
$0(1^{+})$&  &    & 0.3 & 0.7 & 0.9 & 2.2 \\
             &  &    & 0.6 & 10.3 & 1.1 & 11.6 \\
              &  &   & 0.8 & 23.0 & 1.2 & 19.7 \\
$0(2^{+})$& 0.3 & 1.7 &&&  0.4 & 1.5\\
             & 0.5  & 13.6 &&&  0.6 & 11.7 \\
             & 0.6  & 25.4 &&&  0.7 & 22.1 \\
$1(0^{+})$& &   & 1.1  &1.5& $--$ &  $--$ \\
             &  &   & 1.5 &12.3 &$--$ &  $--$ \\
               & &  & 1.7  &24.4 &$--$ &  $--$ \\
$1(1^{+})$& 1.5  & 0.8 & &   &$--$ &  $--$\\
             & 2.0  & 13.5 &  &   &$--$ &  $--$\\
             & 2.2  & 21.1 &  &  &$--$ &  $--$ \\
$1(2^{+})$&  &    & 3.5  &0.4&$1.2$ &  $0.7$\\
             &  &   & 4.4  &13.8&$1.4$ &  $5.5$ \\
             &  &   & 4.7  &21.3 &$1.6$ &  $19.1$ \\
\toprule[2pt]
\end{tabular}
\end{center}

\end{table}

Since the $D\bar{D}^*/B\bar{B}^*$ states can carry different $G$ parities (and different $C$ parities for isoscalar state), we present the results with different $G$ parities in 2-3$^{rd}$ and 4-5$^{th}$ columns, respectively. For other cases, the $G$ parities are fixed for certain isospin and spin parity, we leave the forbidden states blank in the tables. In the doubly heavy sector, the $G$ parity also does not involve, so we list the results with different isospin, spin and parities $I(J^P)$ in the last two columns of the Tables.  Due to the heavy quark symmetry, the potentials in the bottom sector are analogous to these in the charm sector. From Eq.~(\ref{Eq:L}), one can find that the potential in the charm sector should have the same form as that in the bottom sector. At first glance, the results in Table~\ref{Tab: DD bound state} are analogous to  those in Table~\ref{Tab: BB bound state} as expected. With the same cutoff, the interactions in the bottom sector are stronger than the interactions in the charm sector. It leads to smaller cutoff in the bottom sector required to produce a bound state as shown in tables.

From tables, we can find that the bound states  with quantum numbers $I^G(J^P)=1^+(1^+)$  can be produced from the $D\bar{D}^*/B\bar{B}^*$ and $D^*\bar{D}^*/B^*\bar{B}^*$ interactions at  cutoff about 2 GeV. These four states are usually related to the experimentally observed $Z_c(3900)$, $Z_c(4020)$,  $Z_b(10610)$, and $Z_b(10650)$ in the literature~\cite{Zyla:2020zbs,Chen:2016qju,Guo:2017jvc,Guo:2019twa,Dong:2017gaw}.  An isoscalar sate $0^+(1^+)$ is also produced from the $D\bar{D}^*$ interaction, which carries quantum numbers as the $X(3872)$. Though the studies in the literature do not support the $X(3872)$ as a completely molecular state, considerable $D\bar{D}^*$ component is still important to reproduce its property~\cite{Close:2003sg,Barnes:2003vb,Kalashnikova:2005ui}.  

Besides above states, a state with $0^-(1^+)$ is also produced from the $D\bar{D}^*/B\bar{B}^*$ interaction.  For $D^*\bar{D}^*/B^*\bar{B}^*$ interactions, three isoscalar states with $0^+$, $1^+$, and $2^+$ can be produced, and an isovector state with spin parity $0^+$ is also   produced at a cutoff about 1. The $D^*\bar{D}^*/B^*\bar{B}^*$ state with $1(2^+)$ is produced at a cutoff larger than 4 GeV.   It is interesting to see that an isoscalar scalar state is produced from the $D\bar{D}/B\bar{B}$ interaction, which was also predicted in the chiral unitary approach~\cite{Gamermann:2006nm}. As the $D^*\bar{D}^*/B^*\bar{B}^*$ state with $1(2^+)$, production of the $D\bar{D}/B\bar{B}$ state with $1(0^+)$ also requires considerably large value of the cutoff. 

Now we turn to the doubly heavy sector.   Compared with the hidden heavy sector, fewer states are suggested as shown in the tables. No bound state is produced from $D{D}/\bar{B}\bar{B}$ interaction, and an isoscalar $DD^*/\bar{B}\bar{B}^*$ state with $0(1^+)$ is produced at cutoff about 1 GeV. In the $D^*D^*/\bar{B}^*\bar{B}^*$ interaction, two isosclar states with spin parities $1^+$ and $2^+$ and an isovector state with $2^+$ are produced. 

In Table~\ref{Tab: ZcZb} we collect the results for five states which can be related to the experimentally observed $X(3872)$, $Z_c(3900)$, $Z_c(4020)$, $Z_b(10610)$ and $Z_b(10650)$ and provide more informations for reference. The results for the $B\bar{B}^*$ state with $0^+(1^{+})$ is also presented for comparison.  
\renewcommand\tabcolsep{0.16cm}
\renewcommand{\arraystretch}{1.25}
\begin{table}[hbtp!]
\begin{center}
\caption{The binding energy $E_B$ of some bound states with  selected values of cutoff $\Lambda$. The results for virtual (V) and bound (B) states in full model and these without and only with $J/\psi(\Upsilon)$ exchange are listed $2-3^{rd}$, $4-5^{th}$, $6-7^{th}$, $8-9^{th}$ columns, respectively. Other notations are the same as in Table~\ref{Tab: BB bound state}. 
\label{Tab: ZcZb}
\label{diagrams}}
	\begin{tabular}{c|rrrrrrrr}\bottomrule[2pt]
 & \multicolumn{2}{c}{Full(V)} & \multicolumn{2}{c}{Full(B)}& \multicolumn{2}{c}{\rm No $J/\psi(\Upsilon)$}& \multicolumn{2}{c}{$J/\psi(\Upsilon)$}	 \\\hline
$[I^G(J^{P})]$   &  $\Lambda$ & $E_B$ &  $\Lambda$ & $E_B$&  $\Lambda$ & $E_B$ &  $\Lambda$ & $E_B$ \\\hline
$D\bar{D}^*[0^+(1^{+})]$& $--$ & $--$& 0.5  & 2.3 & 0.5  & 1.5 & $--$ &  $--$ \\
    $X(3872)$         & $--$ & $--$& 0.7  & 13.4 & 0.7 & 7.8 & $--$ &  $--$ \\
             & $\mathit{0.2}$ & $\mathit{2.6}$& 0.8  & 23.3 & 0.8 & 15.3 & $--$ &  $--$ \\
$D\bar{D}^*[1^+(1^{+})]$& $\mathit{1.4}$ & $\mathit{22.8}$&1.9  & 2.4 & $--$ &  $--$ & $2.5$ &  $1.1$ \\
  $Z_c(3900)$        & $\mathit{1.5}$ & $\mathit{8.3}$  &2.0 & 7.1  & $--$ &  $--$  & $2.6$ &  $6.6$ \\
           & $\mathit{1.6}$ & $\mathit{2.3}$  & 2.1 & 17.1 & $--$ &  $--$  & $2.7$ &  $17.3$ \\
 $D^*\bar{D}^*[1^+(1^{+})]$& $\mathit{1.4}$ & $\mathit{23.4}$&2.3  & 2.8 & $--$ &  $--$ &3.6 &1.7\\
   $Z_c(4020)$          & $\mathit{1.6}$ & $\mathit{5.3}$ & 2.6 & 10.2  & $--$ &  $--$ & 3.8 &  6.8\\
         & $\mathit{1.8}$ & $\mathit{0.5}$    & 2.8 & 20.0  & $--$ &  $--$ &4.0 & 18.4\\\bottomrule[1.pt]
$B\bar{B}^*[0^+(1^{+})]$& $--$ & $--$ & 0.3 &1.5 & 0.3 & 0.8 & $--$ &  $--$\\
            & $--$ & $--$  & 0.5  & 11.9 & 0.5 & 10.6 & $--$ &  $--$ \\
            & $\mathit{0.1}$ & $\mathit{4.4}$  & 0.6  & 21.4 & 1.5 & 20.2 & $--$ &  $--$ \\
$B\bar{B}^*[1^+(1^{+})]$& $\mathit{0.9}$ & $\mathit{19.5}$ & 1.6  &2.4& $2.2$ &  $0.8$ & $3.6$ &  $1.2$ \\
  $Z_b(10610)$     & $\mathit{1.0}$ & $\mathit{4.4}$    & 1.9 & 10.9& $2.6$ &  $8.4$ & $3.9$ &  $9.8$ \\
             & $\mathit{1.1}$ & $\mathit{1.2}$& 2.1  &22.7& $3.0$ &  $25.4$  & $4.1$ &  $21.5$ \\
 $B^*\bar{B}^*[1^+(1^{+})]$& $\mathit{0.8}$ & $\mathit{16.3}$& 1.5 & 0.8  & 2.2  &1.2& $4.1$ &  $0.8$ \\
  $Z_b(10650)$  & $\mathit{0.9}$ & $\mathit{7.9}$   &2.0 & 13.5  & 2.7  &8.3 & $4.6$ &  $9.4$ \\
             & $\mathit{1.0}$ & $\mathit{1.4}$& 2.2 & 21.1  & 3.1   &20.4 & $5.0$ &  $23.2$ \\
\toprule[2pt]
\end{tabular}
\end{center}

\end{table}

The cutoffs to produce the $D\bar{D}^*$ and $D^*\bar{D}^*$  states with $1^+(1^+)$ are about 2 GeV. It is  larger than those to produce $0^+(1^+)$ state  and some other states listed in Table~\ref{Tab: DD bound state}, which is consistent with the results in Ref.~\cite{He:2015mja}. The possibility of the $Z_c(3900)$ as a virtual state was suggested by some authors in Refs.~\cite{Guo:2013sya,He:2017lhy}. We present possible virtual states in the $2^{nd}-3^{rd}$ columns.  If we assume that the $Z_c(3900)$ and $Z_c(4020)$ are virtual states,  the gap of the cutoff will reduce.  Such reduction is also found in the hidden-bottom sector. Hence, considering the model uncertainties, a cutoff about 1 GeV is reasonable to estimate the possibility of existence of molecular states from the interactions in the current work.  With such criterion,  the results in Table~\ref{Tab: DD bound state} and Table~\ref{Tab: BB bound state} suggest that the $\mathcal{P}\bar{\mathcal{P}}^*[1^-(0^+)]$ and $\mathcal{P}^*\bar{\mathcal{P}}^*[1^-(2^+)]$ states are disfavored and the bindings of states $\mathcal{P}^*\bar{\mathcal{P}}[1^+(1^+)]$, $\mathcal{P}^*\bar{\mathcal{P}}^*[1^-(0^+), 1^+(1^+)]$, $B\bar{B}^*[0^-(1^+)]$,  and $\mathcal{P}^*{\mathcal{P}}^*[1^+(2^+)]$ are weaker than other states listed. 

In the current work, we consider the heavy meson exchange as in  Refs.~\cite{Aceti:2014uea,He:2015mja}.  In Table~\ref{Tab: ZcZb}, we give the results without and only with $J/\psi(\Upsilon)$ exchange of the six states as example.  Without the $J/\psi$ exchange the two hidden-charm states with $1^+(1^+)$ can not be produced while without the $\Upsilon$ exchange the two hidden-bottom states are still found with a little larger cutoff.  Only with the  $J/\psi(\Upsilon)$ exchange,  the bound state can be found in both cases while the hidden-bottom states requires larger cutoffs. It suggests that the $J/\psi$ exchange in the charm sector is more important than the $\Upsilon$ exchange in the bottom sector due to the larger mass of  $\Upsilon$ meson.  For the $0^+(1^+)$ state,  because the sign of isospin factor ($c$ in Table~\ref{flavor factor}) changes, no bound state can be produced only with $J/\psi(\Upsilon)$ exchange. 

\section{Summary} \label{Sec: Summary}
 
 In the current work, we preform a systematic study about the molecular states produced from the hidden heavy and doubly heavy systems.  The potentials are constructed with the help of the Lagrangians under heavy quark and chiral symmetries, which lead to the similarity of the results in charm and bottom sectors. With the potential, the scattering amplitude is obtained within the qBSE approach, and the molecular states can be searched for as the pole of the scattering amplitude. 
The experimentally observed $Z_c(3900)$, $Z_c(4020)$,  $Z_b(10610)$, and $Z_b(10650)$ can be reproduced with quantum numbers $I^G(J^P)=1^+(1^+)$.  

The observation of the $P_c(4312)$ is a surprise in the molecular state picture because it can not be produced with only $\pi$ exchange. The vector exchange should be dominant in its formation as a molecular state. In the current work, the vector exchanges, including the heavy meson $J/\psi(\Upsilon)$, are introduced to mediate the interaction. The results suggest that there exists an isoscalar scalar bound state produced from the $D\bar{D}/B\bar{B}$ interaction. Such state has been suggested in the chiral unitary approach~\cite{Gamermann:2006nm}. However, an experimental measurement  at Belle can not observe a structure corresponding to such state~\cite{Chilikin:2017evr}.  More theoretical analyses were done in Ref.~\cite{Wang:2019evy,Wang:2020elp,Dai:2020yfu}, which suggested more experimental efforts to clarify the existence of such state. For the $D^*\bar{D}^*/B^*\bar{B}^*$ interaction, besides the states corresponding to the $Z_c(4020)$ and $Z_b(10650)$ the isoscalar states with spin parities $0^+, 1^+, 2^+$ and isovector state with $0^+$ are also suggested by the calculation.  

Now that the experimentally observed hidden heavy $XYZ$ particles can be interpreted generally, it is interesting to study the doubly heavy molecular states. After replacing one of the particle by its antiparticle, one can obtain the double heavy system in the same theoretical frame. The results suggest that an isoscalar state with $1^+$ is produced from $DD^*/\bar{B}\bar{B}^*$ interaction, and two isoscalar states with spin parities $1^+$ and $2^+$ and a isovector state with $2^+$ are produced from the $D^*D^*/\bar{B}^*\bar{B}^*$ interaction. Such states can be stuided in future high precision measurement at LHCb.

\vskip 10pt

\noindent {\bf Acknowledgement} This project is supported by the National Natural Science
Foundation of China with Grants No. 11675228.

\end{document}